\documentclass[12pt,preprint]{aastex}
\usepackage{epsfig}

\begin{document}

\newcommand{\gsim}{\hbox{\rlap{$^>$}$_\sim$}}

\title{What we learn from the afterglow of GRB 021211}

\author{Shlomo Dado\altaffilmark{1}, Arnon Dar\altaffilmark{1,2}
and A. De R\'ujula\altaffilmark{2}}

\altaffiltext{1}{Physics Department and Space Research Institute,
Technion, Haifa 32000, Israel} \altaffiltext{2}{Theory Division,
CERN, CH-1211 Geneva 23, Switzerland}


\begin{abstract}

The behaviour of the afterglow (AG) of gamma-ray bursts (GRBs)
directly provides, in the cannonball (CB) model, information about
the environment of their progenitor stars.  The well observed early
temporal decline of the AG of GRB 021211 is precisely the one
predicted in the presence of a progenitor's ``wind'' which resulted
in a density profile $\propto 1/r^2$ around the star.  The subsequent
fast fading ---which makes this GRB ``quasi-dark''---
 is the one anticipated if, further away, the interstellar density is
roughly constant and relatively high.  The CB-model fit to the AG clearly
shows the presence of an associated supernova akin to SN1998bw, and allows
even for the determination of the broad-band spectrum of the host galaxy.
GRB 990123 and GRB 021004, whose AGs were also measured very early, are
also discussed.

\end{abstract}


\section{Introduction}

Massive stars are observed to emit a dense stellar wind, continuously
or in a sequence of eruptions, which presumably intensifies before
their death as core-collapse supernovae (SNe). Indirect evidence
for a large mass loss from the progenitor stars during their
late-time evolution is provided by the absence of H lines (and He
lines) in the spectra of SNe of Type Ib (Ic).  In Type II SNe,
direct evidence for a wind emitted shortly before the explosion is
provided by the subsequent observations of emission by the circumburst
material of narrow optical lines (e.g. Salamanca et al. 1998;
2002; Fassia et al. 2001), radio waves and X rays (for a recent
review, see Chevalier 2003).  All these observations indicate that
the circumstellar density profile is $\rho_w \propto r^{-2}$, as
expected from a quasi-steady stellar wind.

Gamma ray bursts (GRBs), in the cannonball (CB) model, are associated
with the death of massive stars in core collapse SN explosions (Dar
\& De R\'ujula 2000a,b; Dado, Dar \& De R\'ujula 2002a,b).  Clear
indications for this association are the effect of the progenitor's
wind on their early-time afterglow (AG) and the presence of a
supernova in their late-time AG.  In a CB-model analysis, there is
very convincing evidence of a GRB--SN association
(Dado et al. 20002a--e; 2003). The CB-model fingerprints of the massive
wind of the progenitor star  have also been spotted in the AGs,
but never before with unquestionable certainty.

In this paper we fit the broad-band optical afterglow of GRB 021211
in the CB model, and extract information about the circumburst
density profile, the associated SN and the host galaxy. We also
rediscuss GRB 990123 and GRB 021004. In these three cases the AG
was observed in the optical band early enough after the onset of
the burst for the CBs to be still moving through the progenitor's
wind.  The observed early AG decline, roughly $t^{-1.6}$, is the
CB-model expectation\footnote{We have claimed before that this
behaviour ought to be $\sim t^{-2}$. In the case of GRB 990123,
discussed in Dado et al. 2002a, that was due to our then-incomplete
understanding of the AGs' broad-band spectra, which we improved in
Dado et al. 2002b.  For GRB 021004 it was an error, which we
correct here, although it is not relevant in practice (the early
data on this AG are not sufficient to distinguish the two declines).}
for a wind density profile $\propto r^{-2}$.  In the well measured early 
AG of GRB 021211, the agreement between the observations and the
CB-model prediction is extremely satisfactory.

In the CB model we assumed GRBs to be associated with SNe similar
to SN1998bw (Dar and De R\'ujula, 2000a). In subsequent work (Dado
et al. 2002a--e; 2003) we found that, indeed, in all cases of GRBs of
known $z$ in which such a SN was in practice visible (all cases
with $z<1.12$), the CB-model fit to the corresponding AG disclosed
its presence.  GRB 021211, at $z=1.006$, is one more example: the
fit clearly requires such a SN contribution, tentatively reported
by Fruchter et al. (2002) and Testa et al. (2003). The CB model
has so few parameters that we could determine via our fits the
unknown contribution of the host galaxy of GRB 021211 to the
different optical bands. The resulting broad-band spectrum snugly
resembles that of other GRB host galaxies (e.g. Gorosabel et
al. 2003) and star-forming galaxies at a similar $z$ (Fruchter et
al. 2002).

In the CB model ``dark'' GRBs without AGs are not expected.  But
the AGs' decline is expected to be often fairly fast, in particular
if the interstellar medium (ISM)
density is relatively high, making their
detection difficult or, in the bygone days of slow GRB localization,
nearly impossible; see Fig. 6 of Dado et al. (2002a) and its
discussion. With faster detectors, more of these faint, fast-declining
``quasi-dark'' GRBs ought to be found, and GRB 021211 is an example.
The detailed shape of the fast AG decline observed from 20 to 150
minutes, roughly as $1/t$, is well fit with an ISM
whose density, beyond the reaches of the wind, is relatively
high and approximately constant.

\section{GRB 021211}
The gamma ray burst 021211 was detected on December 11.471227 UT,
2002, with the HETE FREGATE, WXM, and SXC instruments. It was a
relatively long-duration ($\geq$ 5.7 s in the 8--40 keV band),
single-pulse GRB with a peak flux of $8\times 10^{-7}$ erg cm$^{-2}$
s$^{-1}$ and a fluence of $\sim 10^{-6}$ erg cm$^{-2}$ (Crew et
al. 2002).  Its early-time optical AG was detected by the robotic
telescopes RAPTOR (Wozniak et al. 2002), KAIT (Li et al. 2003) and
S-LOTIS (Park et al. 2002), 90, 108 and 143 s after the burst,
respectively. A fit to the KAIT first 9 data points, in the interval
of 2.2--6.5 minutes after burst, showed a temporal decay with a
power-law index of $-1.60 \pm 0.02$ and a fit to later data, taken
20 to 150 minutes after the GRB, gave a slope of $-0.96 \pm 0.04$
(Chornock et al. 2002), consistent with the decay seen by Price
and Fox (2002a,b)  who were first to report the detection and the
precise localization of the optical AG.  Follow-up observations in
the optical, NIR and radio bands were reported by Fox et al. (2003)
and by other groups in the GCN.  The redshift of the host Galaxy,
$z=1.006\pm 0.002$, was measured with VLT by Vreeswijk et al. (2002)
and confirmed by Della Valle et al. (2003). A flattening of the
light curve due to the host galaxy and a possible contribution from
an associated SN were observed with HST 7 and 14 days after burst
in the BVIH bands (Fruchter et al. 2002) and with VLT-UT4 between
days 30 and 35 in the R band (Testa et al. 2003).  Well-sampled
optical AG observations have been recently reported by Li et
al. (2003).  The published observational data on the broad-band
optical AG of GRB 021211 are shown in Figs.~\ref{f1} and \ref{f2}.

\section{The CB model}

In the CB model bipolar jets of CBs are launched axially in
core-collapse SNe, with initial Lorentz factors $\gamma_0={\cal{O}}(10^3)$.
The CBs are assumed to be produced in an unstable accretion process,
as in quasars and micro-quasars and, as in SS 433, to be made of
ordinary matter\footnote{Balmer H and He lines (e.g., Eikenberry,
et al. 2001) and the K$\alpha$ line of Fe (Migliari et al. 2002)
 were detected in the mildly relativistic CBs of this $\mu$-quasar.}.
Crossing the SN shell and the progenitor's wind with a large
$\gamma$, the front surface of a CB is collisionally heated to keV
temperatures. The quasi-thermal radiation it emits, when no longer
absorbed by the intervening matter, and boosted and collimated by
its relativistic motion, is a single $\gamma$-ray pulse in a GRB.
The cadence of pulses reflects the chaotic accretion and is not
predictable, but the individual-pulse temporal and spectral properties
are (Dar and De R\'ujula, 2000b; for recent reviews, see De R\'ujula
2002; Dar 2003). The ejected CBs, as observed in $\mu$-quasars,
are assumed to contain a tangled magnetic field.  As they plough
through matter, they gather and magnetically scatter its constituent
protons. The re-emitted protons exert an inward pressure on the
CBs, which counters their expansion and makes them reach an asymptotic
radius $R_{_{CB}}$, in minutes of observer's time (Dado et al. 2002a).
The electrons swept in by the CB in its voyage through the wind
and ISM are Fermi-accelerated in the enclosed magnetic maze and
cooled by synchrotron radiation.  Shortly after the GRB, the optical
radiation from the CB is dominated by synchrotron emission from
these electrons.  So far, the CB model was very successful in
fitting the observed broad-band AGs of all GRBs of known redshift
(Dado et al. 2002a--e, 2003).

\section{The fate of a cannonball}

Let $n_p$ be the baryonic number density of the circumburst
material or the ISM, both dominated by protons.
A spherical CB of radius $R_{_{CB}}$ flying through this material
sees, in its rest system,  an incoming flux of protons
entering it with a total momentum per unit time
$\Pi\simeq n_p\,m_p\,c^2\,\gamma^2\,\pi\,R_{_{CB}}^2$,
with $\gamma=\gamma(t)$ the diminishing CB's Lorentz factor.
If the CB's enclosed magnetic field randomizes these protons
so that they are re-emitted isotropically, the CB's surface is subject
to an inward pressure $P=\Pi/(4\,\pi\,R_{_{CB}}^2)$, which is independent
of $R_{_{CB}}$. In Dado et al. 2002a we have assumed that
this pressure is sustained by the equal and opposite pressure
$B^2/(8\,\pi)$ of the CB's enclosed magnetic field (which is thereby
determined) resulting in an equilibrium situation for a CB of
approximately constant $R_{_{CB}}$. We also assumed that $R_{_{CB}}$,
for a CB originally expanding at a transverse velocity
$\beta_0\,c$, can be estimated as the maximum expansion radius
attained as the pressure $P$ opposes the expansion, that is
$R_{_{CB}}\!\sim\! [3\,N_{_{CB}}\beta_0^2/(2\,\pi\,n_p\,\gamma_0^2)]^{1/3}$,
with $N_{_{CB}}$ the CB's baryon number.

To test the above bold assumptions, we have fitted the model to
all broad-band AGs of GRBs of known $z$, with satisfactory results.
In Dado et al. (2002a) we have also tried a model with continuously
expanding CBs, and found it to be completely inadequate, supporting
the ansatz of CBs with constant radius. The estimated value of
$B\sim 10$ Gauss (for $n_p=10^{-2}$ cm$^{-3}$ and $\gamma_0=10^3$)
is adequate for the description of the data, but our original guess
$R_{_{CB}}={\cal{O}}(10^{14})$ cm turned out to be an overestimate
by at least one order of magnitude (Dado et al. 2002b).  In this
paper we subject the CB's defining property ---that they reach an
approximately constant radius--- to a severe test, by studying in
detail its consequences for very early AGs.

Let $\theta$ be the angle between the direction of a jet
of CBs and the observer.
The Doppler factor of the light the CB emits
is well approximated by
$ \delta(t)\approx 2\,\gamma(t)/ (1+\theta^2\, \gamma(t)^2)$
in the domain of interest for GRBs: large $\gamma$ and small $\theta$
(typical values ensuing from our fits are $\gamma_0\sim 1/\theta\sim 10^3$).
The relation between time ---as measured by the observer--- and distance
---as travelled by the CB--- is:
\begin{equation}
dx = {\gamma\,\delta\over(1 +  z)}\;c\, dt .
\label{dxdt}
\end{equation}
The ambient protons that a CB scatters in the interval $dx$ slow it
down by an amount\footnote{We call $x$ and not $r$
the distance from the progenitor, not to insinuate a hypothesis
of spherical symmetry.}:
\begin{equation}
d\gamma=-{\pi\,R_{_{CB}}^2 n_p\gamma^2\over N_{_{CB}}}\,dx,
\label{dgamma}
\end{equation}
where $N_{_{CB}}$ is the CB's baryon number, for which our reference
value is $6\times 10^{50}$. The function $\gamma(t)$
can be explicitly found by quadrature in the two cases of interest here:
a density profile $n(x)=n_w\,(x_w/x)^2$ (for a ``windy''
neighbourhood) and a constant density (an adequate approximation
as the CBs get further away into the ISM).

For an ISM of constant density, $\gamma(t)$ depends on $\theta$,
on the initial
$\gamma=\gamma_0$ as a CB exits the denser wind domain, and
on  $x_\infty=N_{_{CB}}/(\pi\, n_p\, R_{_{CB}}^2)$, a deceleration
parameter. The value of $\gamma(t)$ is the real root of the cubic:
\begin{equation}
{1\over\gamma^3}-{1\over\gamma_0^3}
+3\,\theta^2\,\left[{1\over\gamma}-{1\over\gamma_0}\right]=
{6\,c\, t\over (1+z)\, x_\infty}\, .
\label{cubic}
\end{equation}
It takes a distance $x_{1/2}=x_\infty/\gamma_0$ for $\gamma(t)$
to descend to $\gamma_0/2$.
Our fitted values of $x_{1/2}$ are in the range 0.1 to 1 kpc, corresponding
to AGs that fade in the observed characteristic times of order  
$(1+z)\,x_{1/2}/(c\, \delta_0)$ $\sim$ days. This range may be
affected by an observational bias: AGs with a smaller $x_\infty$ (e.g. those
with CBs moving in a relatively high density ISM) decay faster in time
and are ``quasi-dark'': harder to detect. GRB 021211 is one such case.

We do not report the function $\gamma(t)$ for the small
values of $t$ corresponding to the CBs crossing the parent
stellar wind, since, for typical parameters, the
fractional energy loss $\Delta\gamma/\gamma$ is negligible
in that interval. The ``canonical'' stellar wind of a very
massive star has a rate $\dot M_w\!\sim\! 10^{-4}\,M_\odot$ y$^{-1}$
and a velocity $v_w\!\sim\!100$ km/s, so that
$\rho(x)\!\approx\!\rho_w\,(x_w/x)^2$ with
$\rho_w\,x_w^2\!\approx\! \dot M_w/(4\pi\,v_w)\!\sim\! 5.1\times 10^{13}$ 
g cm$^{-1}$,
or $n_w\,x_w^2\!\sim\! 3\times 10^{37}$ cm$^{-1}$.
The actual observations of $M_w$ and $v_w$
span an order of magnitude around
these canonical values.
It follows from Eq.~(\ref{dgamma}) that the
condition $\Delta\gamma/\gamma\!\ll\! 1$, for a wind profile extending
from $x_i$ to $x_f\!\gg\! x_i$, is
$R_{_{CB}}^2\!\ll\! N_{_{CB}}x_i/(\pi\,n_w\,x_w^2\,\gamma)$.
For observations starting at $\Delta t=100$ s after burst, $z=1$
and typical $\gamma=\delta=10^3$,
$x_i=c\gamma\delta\,\Delta t/(1+z)=0.5$ pc, and
the constraint is $R_{_{CB}}\!<\!  10^{14}$ cm, which we
have found to be amply satisfied (Dado et al. 2002b).

The fact that $\Delta\gamma/\gamma$ is typically small as the wind is
crossed has, as we shall see anon, an important consequence:
{\it the shape of the early optical AG locally and directly reflects the 
shape of the circumburst density profile}.

\section{The GRB afterglow in the CB model}

The AG ---the persistent radiation in
the direction of an observed GRB--- has three origins: the ejected CBs, the
concomitant SN explosion, and the host galaxy (HG). These components are
usually unresolved in the observed ``GRB afterglows'', so that the
 light curves and spectra are measures of
the corresponding cumulative energy flux density:
$
F_{_{AG}}=F_{_{CBs}}+F_{SN}+F_{_{HG}}.
$
In all observed cases, but one (GRB 021004, discussed in Dado et al.
2003), it is sufficient to approximate the ensemble of CBs in the AG
phase as a single (or dominant) CB.

The contribution $F_{SN}$ of the SNe is approximated by that of
SN1998bw (Galama et al. 1998),
 displaced to the GRB's redshift\footnote{The cosmological
parameters we use are: $\rm H_0=65$ km/(s Mpc), ${\rm \Omega_M}=0.3$
and ${\rm \Omega_\Lambda}=0.7$.}. In the CB model
 the pair GRB 980425 \& SN1998bw is in no way
exceptional (but for the accidentally small $z$ and large $\theta$
of the GRB). Thus, the use of SN1998bw as a candidate GRB-associated
standard candle makes sense.


The optical AG of a CB is dominated by synchrotron radiation from
the ISM electrons that penetrate in it. We argued in Dado et al.
(2002b) that these electrons are Fermi-accelerated in the CB-enclosed
magnetic maze and cooled by synchrotron radiation to a distribution
(in the CB's rest system) $dn_e/dE\propto E^{-2}$ below the {\it
injection bend}  energy at which they enter the CB
($E_b=\gamma(t)\,m_e\,c^2$); and to a distribution
$dn_e/dE\propto E^{-(p+1)}$ above
$E_b$, with $p=2$ in analytical approximations and $p=2.2$ in
numerical simulations.  The emitted synchrotron radiation has a
corresponding double power-law form $\nu\,dn_\gamma/d\nu \propto
\nu^{-\alpha}$, with a power-law index $\alpha_l\approx 0.5$ well
below a {\it bend frequency} $\nu_b$, and $\alpha_h=p/2\approx 1.1$
well above it. Broad-band fits to the data on the AGs of all GRBs
of known $z$ result in $\alpha_l\approx 0.6\pm 0.1$, $\alpha_h\approx
1.1\pm 0.1$. In practice $\alpha_l$ is less well determined than
$\alpha_h$, the CB-model fits being good even for a fixed $\alpha_h=
1.1$. We attribute this to the fact that the extracted $\alpha_h$
is sensitive to the AG light-curve at relatively late times, when
CBs are typically hundreds of parsecs away from the progenitor and
the absorption in the host, which is hard to ascertain, may be
minimal.

The value of $E_b$ and the estimate of the CB's magnetic field
described in the previous section allow us to estimate
the bend frequency of the synchrotron-radiation spectrum:
\begin{equation}
\nu_b(t) \simeq {1.87\times 10^3\, [\gamma(t)]^3\,\delta(t)\over 1+z}\,
\left[{n_p(x)\over 10^{-3}\;{\rm cm}^{-3}}\right]^{1/2}\, \rm Hz.
\label{nubend}
\end{equation}
In the observer frame, this radiation is Doppler-boosted and
collimated by the relativistic motion of the CB, and redshifted
by the cosmic expansion. Its explicit form, extending from
radio to X-ray frequencies, is given in Eqs.~(4,6,8)
of Dado et al. 2002b; here we only report its behaviour near the
optical domain, which is simple above and below the injection bend.
Let $\alpha=(\alpha_l,\,\alpha_h)=(0.5,1.1)$ be the predicted power
indices at $\nu\!\ll\! \nu_b$ and $\nu\!\gg\! \nu_b$, respectively. The 
time and frequency dependence of the energy fluence in these
limits is:
\begin{equation}
F_\nu \propto { A'(\nu,t)\;n_e^{(1+\alpha)/2}\,
[\gamma(t)]^{3\alpha-1}\, [\delta(t)]^{3+\alpha}}  \,
\nu^{-\alpha}\, ,
\label{sync}
\end{equation}
where $n_e=n_e(x)\simeq n_p(x)$ is the electron density along the
CBs' trajectory and $A'(\nu,t)$ corrects for absorption in the
host galaxy and in ours, its possible time dependence originating
in the kiloparsec length of the CBs' trajectory in the host galaxy.
Self-absorption in the CB is irrelevant
at the optical and NIR wavelengths relevant here.
In the CB model the early AG is dominated by the last significant
pulse (or CB) of the GRB, so that $t$ in Eq.~(\ref{sync}) is the time
after that pulse. The nuance may be significant at very small $t$.

In deriving Eq.~(\ref{sync}) we have assumed that a
fixed fraction of the energy-deposition rate by ISM electrons in a CB
($\pi\,R_{_{CB}}^2\,n_e\,m_e\,c^3\,\gamma^2$, in the CB's
system) is re-emitted as the AG. The AG spectrum is a
function of $\nu$ and $\nu_b$, so that at fixed $\nu$ it depends on
$n_e=n_p$ via $\nu_b$, as in Eq.~(\ref{nubend}). Thus a result which is
not linear in $n_e$, and peculiar at first sight.

In our fits we assume as the circumburst density profile
a constant plus a ``windy''  term:
\begin{equation}
n(x)=n_0+n_w(x_w/x)^2\equiv n_0\,[1+(\bar x/x)^2].
\label{density}
\end{equation}

\section{The very early optical AG}

For typical parameters,
the first few parsecs of a CB's voyage ---as it crosses the ``wind''
material emitted by the progenitor star--- are seen by an observer
in the first few minutes of the AG. During that ``early'' time, the
ambient density is high enough for $\nu_b$, as in Eq.~(\ref{nubend}),
to be comfortably above the optical frequencies, so that
$\alpha=\alpha_l=0.5$ in Eq.~(\ref{sync}).
We have seen that, at ``early'' times, a CB
 has its Lorentz factor insignificantly changed
by collisions with the wind material, so that the observer's time and
the CB's travelled distance, as in Eq.~(\ref{dxdt}),
 are strictly proportional. Thus Eqs.~(\ref{sync},\ref{density}) collapse to:
\begin{equation}
F_\nu\simeq n_e^{(1+\alpha)/2}\propto
\left[1+{\left(\bar t/ t\right)^2}\right]^{3/4}.
\label{early}
\end{equation}
Had we used the observed $\alpha_l=0.6\pm 0.1$, as opposed to the
naive theoretical $\alpha_l=0.5$, we would have obtained
$F_\nu\simeq t^{-1.6}$ at veary early times, in close agreement
with the fit by Chornock et al. (2002) for GRB 021211,
and the observations of Akerlof et al. (1999) for
GRB 990123 and Fox et al. (2002) for GRB 021004. But we shall
see that Eq.~(\ref{early}) provides an excellent fit to the data.
The observed deviations relative to the smooth curve
of Eq.~(\ref{early}) reflect the variations of the wind
that generated the profile. In the case of GRB 021211,
these variations are at the $\sim 10$\% level (Li et al. 2003).

\section{The overall AG of GRB 021211 in the CB model}

\begin{figure}[t]
\plotone{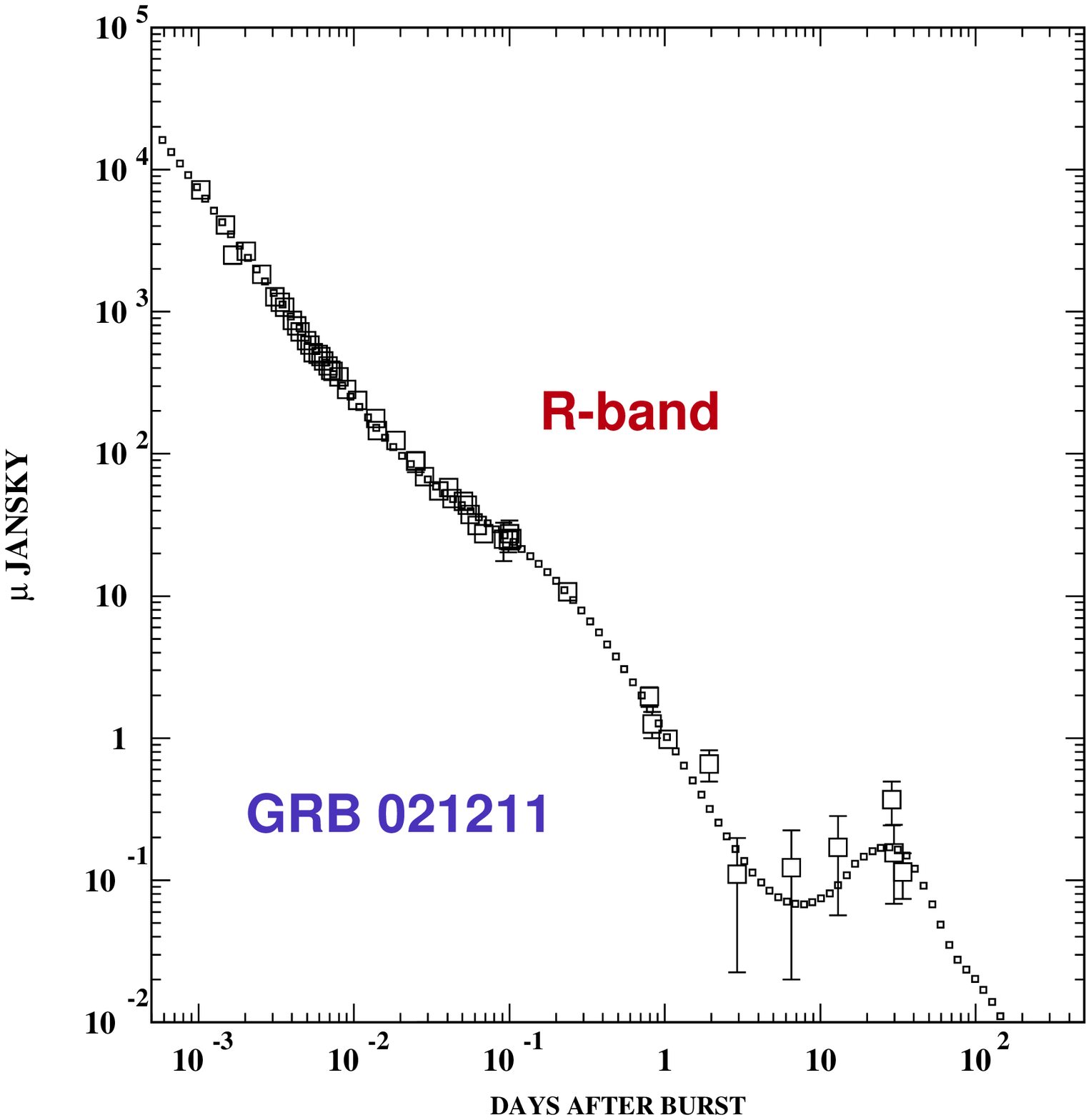}
\figcaption{The optical
observations of the R-band
AG of GRB 021211, and their CB-model fit.  The ISM density
is a constant plus a ``wind'' contribution decreasing as the inverse square
of the distance.  The two contributions are equal at $\bar x\simeq 1.2$ pc, a
distance reached by the CBs in an observer's time $\bar t\simeq 0.025$ days
after burst. The data are those
reported to date, in the GCN notices quoted in section 2, in  Fox et al. 
2003 and in Li et al. 2003.   The
contribution of a SN1998bw-like SN at the GRB position
is discernible at late times.
The host galaxy's contribution, which was fitted, is subtracted in this plot.
\label{f1}}
\end{figure}

\begin{figure}[t]
\plotone{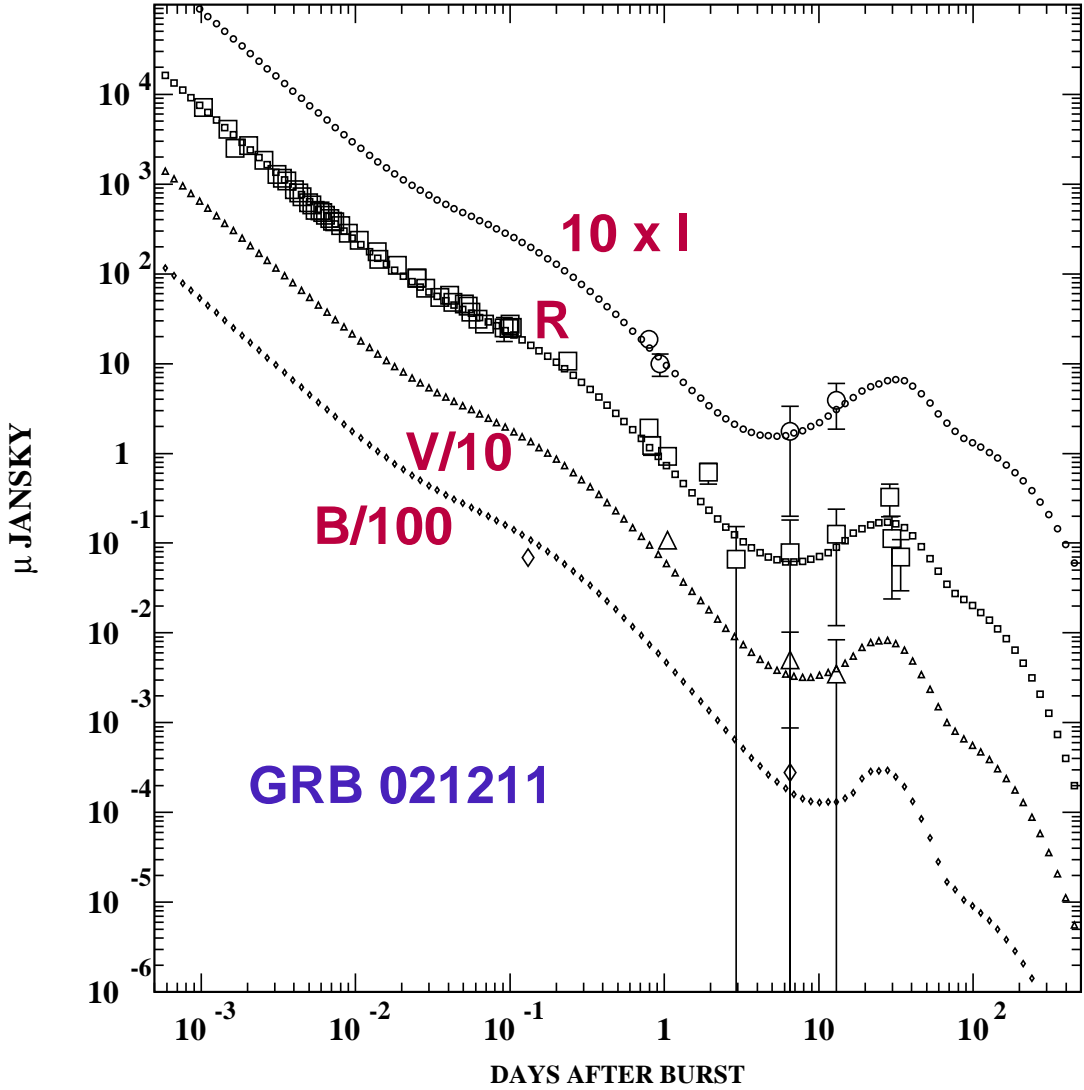}
\figcaption{The wide-band CB-model fit to the optical and NIR observations
for GRB 021211.
For clarity, only the better-sampled I, R, V and B bands are shown,
scaled by 10, 1, 1/10 and 1/100, respectively. The
contribution of a SN1998bw-like SN at the GRB position
is discernible at late times.
The host galaxy's contributions, which were fitted, are subtracted in  
this plot.
\label{f2}}
\end{figure}

The copious early data on GRB 021211 obtained by Li et al. (2003),
along with all other data communicated in the GCN notices
quoted in section 2 and in
Fox et al. 2003 are shown in Fig.~\ref{f1} for the R-band,
and in Fig.~\ref{f2} for the broad-band optical (IRVB) and NIR
(HKJ) passbands. In fitting these results we neglect the
 unknown extinction in the host, due to the lack of relevant spectral
information\footnote{An unabsorbed SN1998bw-like contribution fits the
late-time broad-band data, indicating that host extinction is not large.}.
For this GRB the R-band data are relatively so copious that it is useful
to perform separate R-band and broad-band fits.

In our fit, the density is that of Eq.~(\ref{density}) and the
limiting exponents in the synchrotron spectrum
$\propto\nu^{-\alpha}$ are set to their theoretical values
$(\alpha_l,\alpha_h)=(0.5,1.1)$. The fit is made with
the complete analytical broad-band predictions Eqs.~(4,6,8) of Dado et 
al. 2002b, of which Eq.~(\ref{sync}) are the limiting behaviours and
Eq.~(\ref{cubic}) is the deceleration law.
The best fitted parameters for the R-band fit shown in Fig.~\ref{f1}
are $\gamma_0=262$, $\theta= 1.76$ mrad (i.e.
$\delta_0=431$) and $x_{\infty}=4.7$ kpc, with 51 d.o.f.~and
$\chi^2/{\rm (d.o.f.)}=0.97$ if the out-lying point at $t\!\sim\! 2$ days is 
eliminated. The constant contribution
of the host galaxy was left as a free parameter, for which the
fit returned $F_{_{HG}}=0.34\, \mu$Jy.
The contribution of a standard-candle SN1998bw
at the GRB position is discernible in Fig.~\ref{f1}, in which $F_{_{HG}}$
has been subtracted.

We have also tried a
fit with an arbitrary power $a$ in the windy term: $n_w(x_w/x)^a$,
resulting in a best fitted $a=1.92\pm 0.03$, close to the
``canonical'' expectation $a=2$. Thus, the agreement of the fits
with the canonical ``windy'' power is not a consequence of the
scarcity of data. Also, at early times, the fit is totally
insensitive to the other parameters ($\gamma_0$, $\theta$ and $x_\infty$).
Thus, the result for $a$ is robust: the signature of a wind is there.

The overall normalization of the AG is also a free parameter in our
fits; it is proportional to the ambient density $n(x)$, to the CBs' cross
sections, to their number, and to the unknown fraction of the
energy of the gathered ISM electrons that is re-emitted as the AG.
Thus, extracting a value of $n(x)$ from the fit normalization requires
extra assumptions. Yet, the value of $n(x)$ at some reference distance
(e.g. $n_0$, its limiting value at large distance)
can be extracted from the fact that we use it as a free
parameter in the expression of Eq.~(\ref{nubend}) for $\nu_b$.
The fitted result is $n_0=(2.97\pm 0.22)$ cm$^{-3}$.
Combined with the fitted value
$\bar x=1.2$ pc in Eq.(\ref{density}), this
yields  $\rho_w\,x_w^2=m_p\,n_w\,x_w^2=
(6.8\pm 0.5)\times 10^{13}$ g cm$^{-1}$,
compatible with the expectation, quoted in section 4, for a
``canonical wind'': $5.1\times 10^{13}$ g cm$^{-1}$.

The wide-band fit is shown in Fig.~\ref{f2}; it has essentially
the same fit parameters as the R-band fit.  The contributions
$F_{_{HG}}$ of the host galaxy in the different bands, resulting
from the fit, are 0.50, 0.34, 0.23 and 0.23 $\mu$Jy for the IRV
and B bands, respectively. This extracted spectrum, and in particular
its flattening above the V frequencies, is compatible with that of
other GRB hosts (e.g. Gorosabel et al. 2003) and star-forming
galaxies at $z\!\sim\! 1$  (e.g. Fruchter et al. 2002).

The comparison we have described between the CB-model expectations
and the data for GRB 021211 is simply spectacular. The fits are
good, particularly that to the shape of the R-band light curves.
The agreement between the expected and extracted ambient  density
profiles is eery. A  word of caution, however, is necessary. The
number of parameters in our fits is high: the ``ambient'' parameters
describing the host galaxy ($F_{_{HG}}$) and the density profile
($n_0$ and $\bar x$), the observer's viewing angle $\theta$, and
the 3 quantities that are specific to the CBs:  the overall AG
normalization, $\gamma_0$ and $x_\infty$. Even if the fitted ambient
parameters turn out to have the expected values, a model with this
many parameters has to be very wrong not to describe a very simple
surface: the fluence as a function of frequency and time\footnote{It
is possible to construct models that disagree with observation or
are unphysical, but have even more parameters.  For the case of
GRB 991208, see section 2 of Dado et al. 2002e.}.  Moreover, the
data are not cross-calibrated, their systematic errors are unknown,
the model is no doubt a simplification of a very complicated
phenomenon, and the parameters are not uncorrelated (except at the
early times of particular interest here).  As a consequence of all
this, the ``formal errors'' of the parameters (as given by the
standard program MINUIT) are no doubt underestimated. Whence our
habit of not always making the errors explicit.

\section{The radio AG of GRB 021211}

In the CB model the radio AG at early time is suppressed because of limb
darkening and the finite time it takes the electrons gathered by a CB to
cool radiatively and reach an energy at which they emit synchrotron
radiation of a given typical frequency (Dado et al. 2002b). At later times
the radio emission, like the optical and X-ray emission, becomes
proportional to a high power of the Lorentz factor. 
In the case of the radio AG of GRB 021211, only upper limits (or
perhaps marginal detections) have been reported, by Hoge et al. (2002)
and Fox et al. (2002). The later authors, who analyze the AG in detail,
consider the absence of a stronger radio signal to be a problem. 
In the CB model, it is not.

The wide-band spectrum of AGs is, in the CB model, very simple.
Its only feature ---besides the bend frequency that we have discussed---
is due to self-absorption in the CB, and it is described by a single
parameter $\nu_a$ (Dado et al. 2002b). In the case of GRB 021211
we cannot fix this parameter, having no single secure radio signal.
The best we can do is to show, by way of example, what the radio
signal would be if the value of $\nu_a$ was similar to the average 
for other radio AGs ($\bar \nu_a\!\sim\! 1$ GHz) or to the value for
GRB 021004 (0.98 GHz), which is otherwise
akin to GRB 021211, and is very well described by the CB model
(Dado et al. 2003).  The result of this exercise is shown in
Fig.~\ref{f4}, and it is compatible with the observational limits or 
marginal detections: the largest  ``signals'' 
reported in Fox et al. 2002 are $45\pm 23$ and $60\pm 38$ $\mu$Jy, 
at $t\!\sim\! 5,\,10$ days, both at 8.46 GHz, while Hoge et al. (2002)
obtain a 3$\sigma$ upper limit of 7.5 mJy at 347 GHz, on day $t\!\sim\! 1$.

\section{The GRB proper}

In the CB model, the fluence $F$
from a GRB viewed at a small $\theta$, is amplified by
a huge factor $\delta_0^3$, due to Doppler boosting and
relativistic collimation (Dar \& De R\'ujula 2000):
\begin{equation}
F={(1+z) \,\delta_0^3\over 4\,\pi\, D_L^2}\, E_\gamma\, ,
\label{fluence}
\end{equation}
where $E_\gamma$ is the  total energy in photons emitted by CBs in their 
rest system. The total ``equivalent spherical'', or would-be 
isotropic energy, $E^{iso}$, inferred from the observed fluence, 
is a factor $(\delta_0)^3$ larger than $E_\gamma$. 
In Dado et al. 2002a we deduced that the $E_\gamma$ values
of the GRBs of known $z$ span the surprisingly
narrow\footnote{GRBs in the CB model are much better standard candles
than in the standard model (Frail et al. 2001).}  interval
$10^{44\pm 0.3}$ erg,  the spread in  $E^{iso}$ 
being mainly due to the spread in their values of $\delta_0$
(deduced from the fits to their AGs). For GRB 021211,
the CB-model expectation is $E^{iso} \approx \delta_0^3\,
E_\gamma\approx 0.8\times 10^{52\pm 0.3}$ erg,
in agreement with the observed $E^{iso}\approx 1\times 10^{52}$
erg,  deduced from its measured redshift ($z=1.006$;
Vreeswijk et al. 2002) and fluence
in the 8--40  keV band ($\sim  10^{-6}$ erg  cm$^{-2}$; Crew et 
al. 2002).

\section {GRBs 990123 and 021004}

The AG of these two GRBs was also observed particularly early, by
Akerlof et al. (1999) in the case of GRB 990123.  We have recalibrated
their data assuming $V-R=0.28$ mag (our predicted $\alpha_l=0.5$),
rather than a constant colour.  We have also recalibrated the
later-time data compiled in Castro-Tirado et al. 1999; Kulkarni et
al. 1999; Galama et al. 1999 and Fruchter et al. 1999, assuming
$R-r=0.015$ mag ($\alpha_l=0.5$). The R-band results of our broad-band
fit are shown in Fig.~\ref{f4}, in which $t=0$ corresponds to the
start of the last (second) prominent
pulse in the GRB, 37s after trigger (Briggs et al. 1999).  There is an 
excellent agreement between the early
data tracing the circumburst windy density and
the prediction of Eq.~(\ref{density}). A fit with 
an arbitrary power $a$ in the windy term, $n_w(x_w/x)^a$,
results in a best-fitted $a=1.98\pm 0.06$, again in agreement
 with the ``canonical'' expectation: $a=2$.

Though the data have been recalibrated, the parameters we fit to
the AG of GRB
990123 are close to those in Dado et al. 2002a,b. In particular
$\gamma_0=1204$ and $\delta_0=1630$ ---determined from the {\it
shape} of the AG--- are the highest for any GRB of known $z$,
implying that the {\it magnitude} of the GRB and AG fluences should
also be record-breaking, as they are (even for this GRB, the value of
$E_\gamma$ in Eq.~(\ref{fluence}) is not exceptional).  Since
$\gamma$ and $\delta$ are so large, the bend frequency $\nu_b$ of
Eq.~(\ref{nubend}) ``crosses'' the optical band very late in time,
where the data are quite imprecise.  As a consequence, the R-band
and wide-band fits are very insensitive to $\nu_b$, and we cannot
reliably use its value to determine the absolute magnitude of the
ambient and wind densities. If we assume that the density
profile $n_w\propto x_w^{-2}$  prevails until $n_w$
declines to the typical
superbubble or halo value, $n_p\sim 10^{-3}$, we find that
$\rho_w\, x_w^2=m_p\, n_p\, x_w^2\approx 4.2\times 10^{13}$ g
cm$^{-1}$, again in agreement with the  
expectation, quoted in section 4, for a
``canonical wind'': $5.1\times 10^{13}$ g cm$^{-1}$.
The windy profile would in this case
extend all the way to $x_w\sim 50$ pc, yielding a
total  mass loss of $\sim 40\, M_\odot$ during the life of the very
massive progenitor star. Such a large total mass loss is
not unusual for the progenitors of SNe of Type Ib and Ic.

The CB-model fit to the R-band AG of GRB 021004 is shown in Fig.~\ref{f5}.  
The data are those measured and compiled by Fox et al. 2002; Holland et
al. 2002; Pandey et al. 2002 and Bersier et al. 2003. Once more, the
agreement between the prediction of Eq.~(\ref{density}) and the early data
is good, though these data are neither as abundant nor as close to the GRB
trigger as for the other two cases we discussed, implying that we cannot
claim to have clearly spotted, for this GRB, a typical windy profile.  
The fitted result for the asymptotic density is $n_0=(4.08 \pm 0.17)
\times 10^{-3}$ cm$^{-3}$. Combined with the fitted value $\bar x\sim 5.1$
pc in Eq.~(\ref{density}), this yields $\rho_w\,x_w^2=m_p\,n_w\,x_w^2\sim
1.7\times 10^{12}$ g cm$^{-1}$, a bit over an order of magnitude below
the central ``canonical'' expectation.
 
\section{Discussion and conclusions}

It is, as usual, interesting to compare the CB model with the standard
fireball or blast-wave models of GRBs. For GRB 021211 a detailed SM
analysis is given in Fox et al. 2003. The relevant issues at hand concern
the signatures for ``windy'' neighbourhoods, and the predictive power for
early AGs\footnote{For Fox et al. (2003) the non-observation of a radio AG
is also significant, suggesting that ``the burst may have suffered
substantial radiative corrections''. No such specific cure is necessary in
the CB model.}.

In the standard model (SM)  the scarcity of AGs indicating a windy
circumburst density distribution is a problem, admitted even by its
staunchest defenders (Piran 2001; Price et al. 2002). In the CB model all
the observed AGs that should show a windy signature do. Indeed, the
relation Eq.~(\ref{dxdt}) between observer's time and CBs travelled
distance implies that, for the typical $\gamma_0\sim \delta_0\sim 10^3$,
the AG of a GRB at a cosmological distance must be ``caught'' within the
first few minutes for the CBs to be still travelling close enough to the
parent star, so that the contribution of its wind to the ambient density
is still significant and observable. Only the three GRBs that we have
discussed pertain to this category.

In the SM the fast-decreasing ``early'' optical AG, as first observed in
GRB 990123, is generally attributed to a {\it reverse} shock (M\'esz\'aros
and Rees 1999).  For GRB 021211, this is the interpretation spoused by the
observers (Fox et al. 2003; Li et al. 2003). In this view the index of the
approximate power law of the decay of the early AG is not fixed, and its
normalization is unrelated to that of the late AG.  In the CB model,
contrariwise, the early optical AG does not have its own separate origin:
it is made by the very same mechanism as the late AG. The temporal shape
of the AG is not a succession of power laws with breaks between different
mechanisms or regimes. This temporal shape is not arbitrary: at early
times it is directly related to the shape of the circumburst density
profile, as in Eq.~(\ref{early}), and this relation is fully vindicated by
the data for the expected windy profile.  The AGs shown in our figures
depend, in the ``early'' domain, on only one parameter: the absolute
normalization. The ambient density that defines the early temporal shape
can also be extracted from the CB-model fit, via the bend frequency of
Eq.~(\ref{nubend}), and its magnitude turns out to coincide with the
expected density for the canonical wind of a heavy star.

Fast-declining optical light curves ---masquerading as ``dark'' AGs--- are
expected in the CB model (Dado et al. 2002a), and GRB 021211 is an
example. The CB-model fit to its AG has allowed us to infer the presence
of an associated SN akin to SN1998bw, as for all other GRBs where the SN
could in practice be detected. The fit also yields the ``colour'' of the
host galaxy, which is the expected one.

We have seen how, concerning the early optical AGs, the CB model
is very predictive and successful.  In previous works we have shown
this to be the case also for all properties of the AGs of all GRBs
of known redshift: they can {\bf all} be described in a simple,
analytical and parameter-thrifty manner, which does not involve
multiple choices, multiple mechanisms and multiple exceptions. To
put it in a way with which even the most allegiant apologists of
the SM might agree: the contrast between the CB model and the
standard model is striking.

{\bf Acknowledgement:} This research was supported in part by the Helen
Asher Space Research Fund and by the VPR fund for research at the
Technion. One of us, Arnon Dar, is grateful for hospitality
at the CERN Theory Division.

{}

\clearpage

\begin{figure}[t]
\plotone{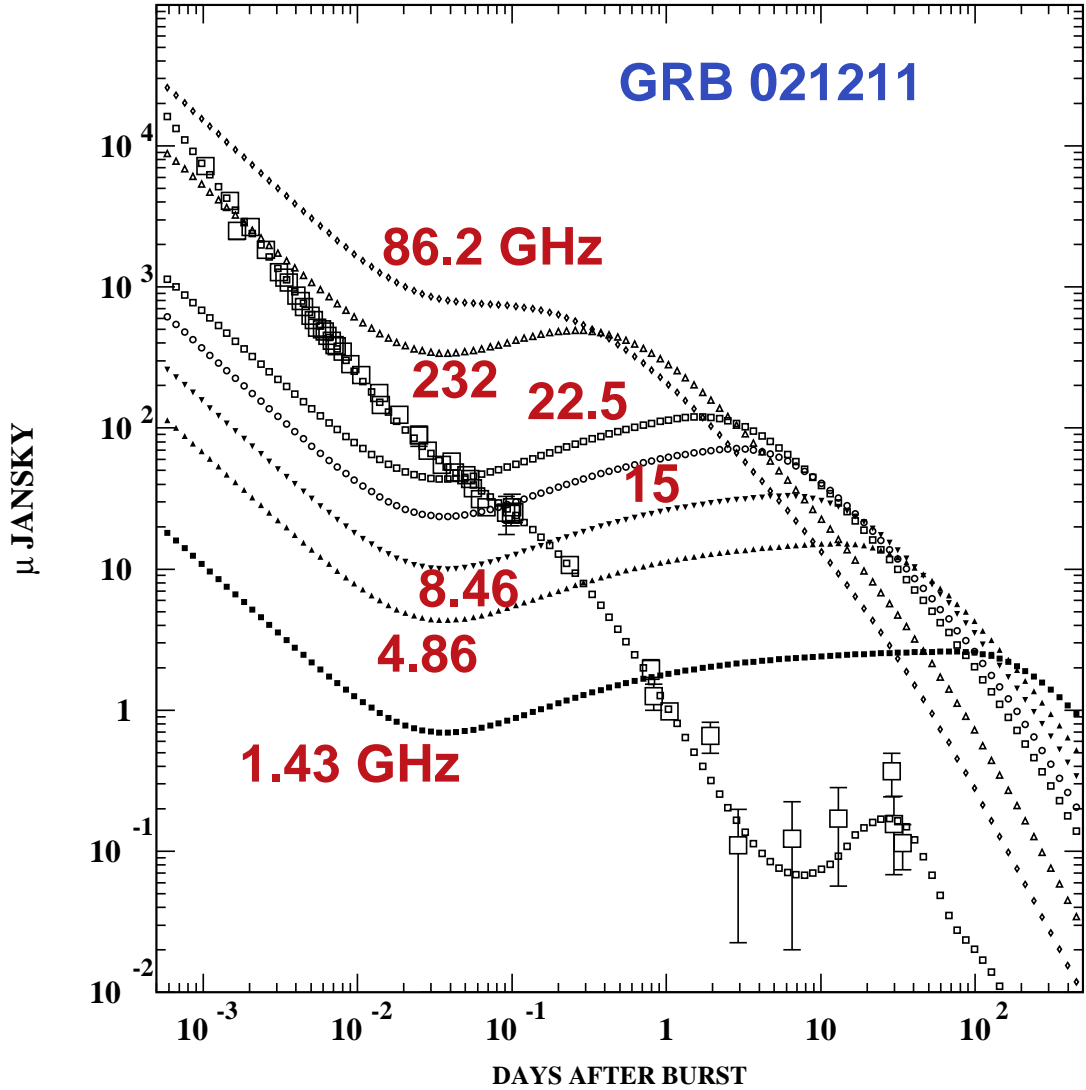}
\figcaption{The optical R-band
AG of GRB 021211, superimposed on the predictions for the radio
AG at the various labelled frequencies, for an assumed $\nu_a=1$ GHz.
\label{f3}}     
\end{figure}

\begin{figure}[t]
\plotone{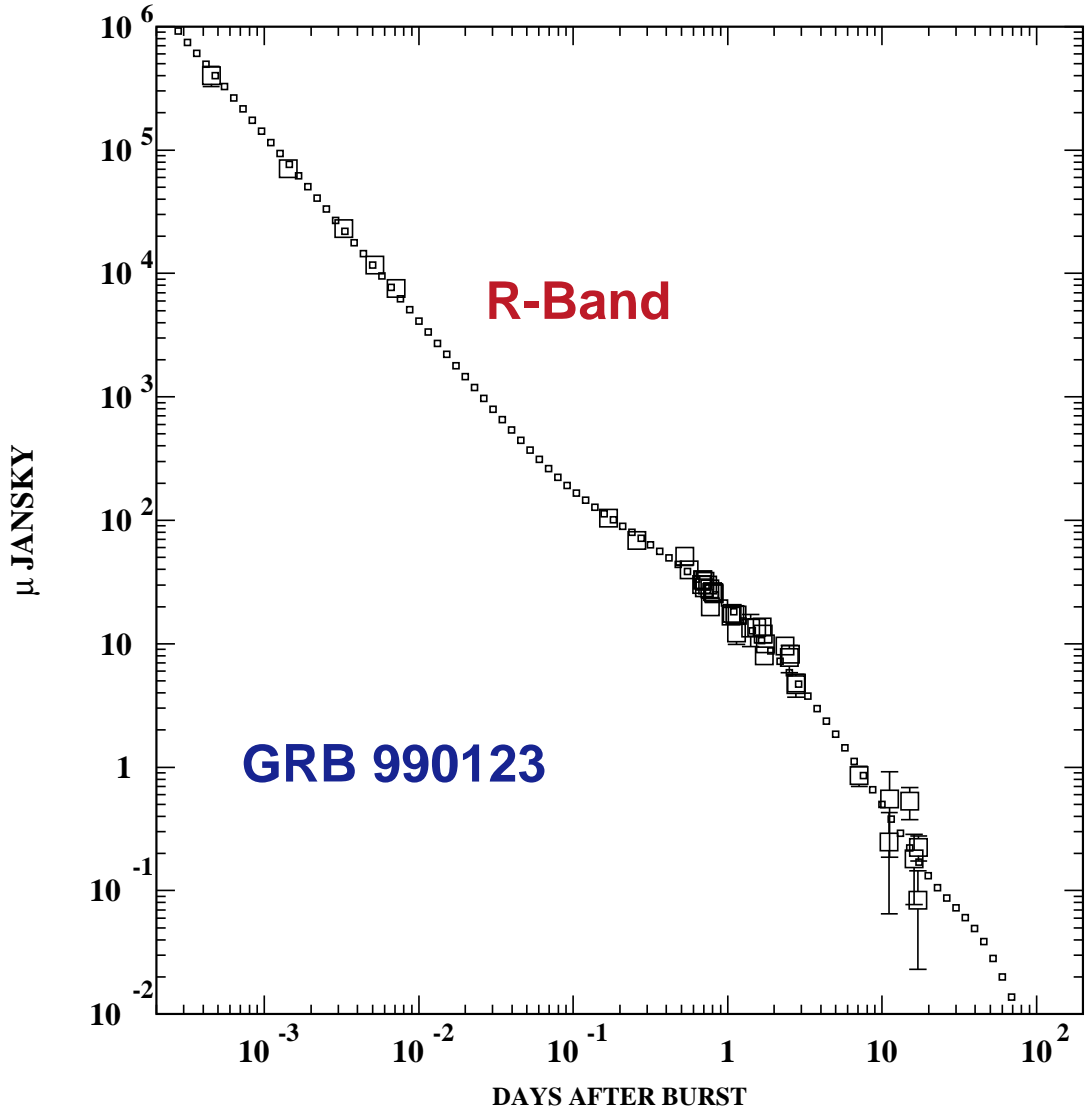}
\figcaption{The optical
observations of the R-band
AG of GRB 990123, and their  CB-model fit. The ISM density
is a constant plus a ``wind'' contribution decreasing as the inverse 
square of the distance. The contribution of the host galaxy has been
subtracted. The small extinction in the Galaxy (Schlegel et al. 1998)
was neglected. 
\label{f4}}     
\end{figure}

\begin{figure}[t]
\plotone{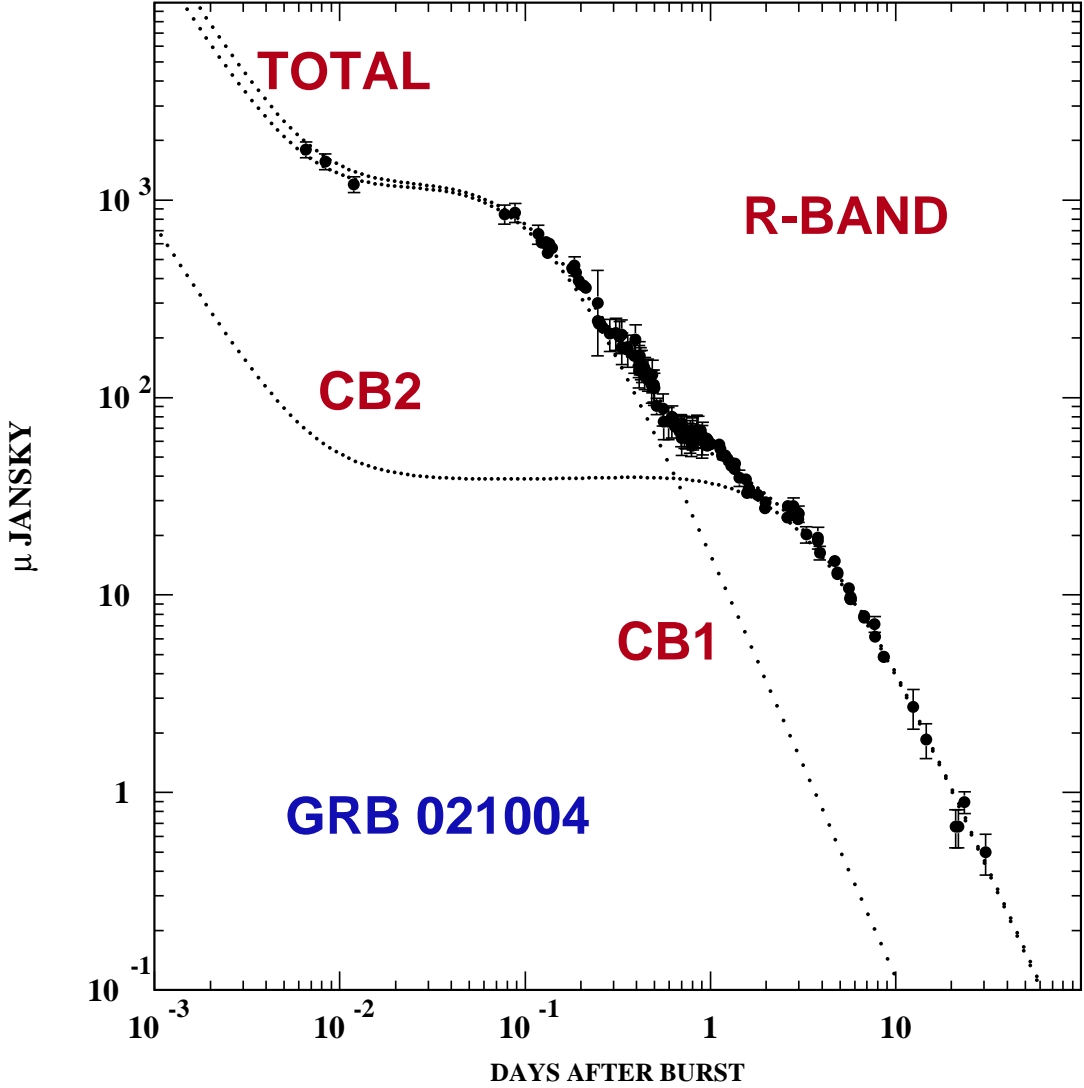}
\figcaption{The optical observations of the R-band AG of GRB 021004,
and their  CB-model fit with two CBs, whose individual contributions
are depicted along with the total (Dado et al. 2002e).  The ISM
density is a constant plus a ``wind'' contribution decreasing as
the inverse square of the distance.  These two contributions are
equal at $\bar x\simeq 3$ pc, a distance reached by the CBs in an
observer's time $\bar t\simeq 0.01$ day after burst. The data are
the same as the ones quoted in Dado et al. 2002e.  The contribution
of a SN1998bw-like SN at the GRB position and the host galaxy's
contribution were subtracted in this plot.
\label{f5}}     
\end{figure}

\end{document}